\documentclass[a4paper,12pt]{article}
\usepackage{amsmath,amsfonts,amssymb,mathrsfs}

\def\diag{\mathop{\rm diag}\nolimits}
\makeatletter
\newcommand{\ps@persona}{%
\renewcommand{\@oddhead}{\hfil{\makebox{\parbox{3cm}{UT-HET 088}}}}
\renewcommand{\@evenhead}{}
\renewcommand{\@oddfoot}{}
\renewcommand{\@evenfoot}{\hfil\rm{\thepage}\hfil}
}
\renewcommand{\thefootnote}{\fnsymbol{footnote}}
\makeatother
\begin{document}
\begin{titlepage}
\vspace{1em}
\begin{center}
\Large{\bf An exact representation isotropic and anisotropic noncommutative phase spaces, and their relations}
\\
\vspace{2em}
{H.~Kakuhata\footnote[2]{hiroshik@eng.u-toyama.ac.jp} and M.~Nakamura\footnote[3]{makoto@jodo.sci.u-toyama.ac.jp}}
\\
\vspace{2em}
$^\dagger${\it\small Faculty of Engineering, University of Toyama , \\
3190 Gofuku, Toyama 930-8555, Japan}\\
$^\ddagger${\it\small Faculty of Science, University of Toyama , \\
3190 Gofuku, Toyama 930-8555, Japan}\\

\end{center}
\begin{abstract}
Noncommutative phase space of an arbitrary dimension is considered.
The both of operators  coordinates and momenta in noncommutative phase space may be noncommutative.
In this paper, we introduce momentum-momentum noncommutativity in addition to coordinate-coordinate noncommutativity.
We find an exact form for the linear coordinate transformation which
relates a noncommutative phase space to the corresponding ordinary one.
As an example, the Hamiltonian of a three-dimensional harmonic oscillator is examined.
\end{abstract}
\thispagestyle{persona}
\end{titlepage}
\renewcommand{\thefootnote}{\arabic{footnote}}
\newpage

\section{Introduction}
 Noncommutativity has become a vital field of research owing to its development in string theories, quantum field theories and quantum mechanics \cite{sewi}.
The open string end points are noncommutative in the presence of the background NS-NS B-field which indicates that the coordinates of D-branes are noncommutative \cite{chho1}. 
There has been a lot of works based on perturbative and non-perturbative field theories in noncommutative space \cite{hike}. 
An extensive research has also been done on noncommutative
quantum mechanical systems \cite{chda,duja,mopo,horv}.

Quantum mechanics in the noncommutative configuration space has attracted much attention.
The noncommutative configuration space is characterized by the commutation relations between $N$-dimensional coordinates $q_i$ and momenta $p_i$
as
\begin{align}
\begin{aligned}
&[q_i,~q_j]=i\theta_{ij},\qquad [q_i~,p_j]=i\delta_{ij},\qquad [p_i,~p_j]=0,\cr
\end{aligned}
\label{eqn:i1}
\end{align}
where, $\theta$ is an $N\times N$ antisymmetric matrix.
The noncommutative quantities $q_i$ and $p_i$ are known to be expressed by the use of the commutative phase space variables $Q_i$ and $P_i$ as follows:\cite{Mezincescu,Deri1,Deri2,Girotti}
\begin{align}
\left\{
\begin{aligned}
&q_i=Q_i-\sum_{j=1}^N\frac{1}{2}\theta_{ij}P_j,\cr
&p_i=P_i,\cr
\end{aligned}
\right.
i=1,\dots,N
\label{i2}
\end{align}
with
\begin{align}
[Q_i,~Q_j]=0,\qquad [Q_i,~P_j]=i\delta_{ij}, \qquad [P_i,~P_j]=0.
\label{i3}
\end{align}
Therefore, the Hamiltonian in the noncommutative phase space is rewritten as
\begin{align}
H(q,~p)=H(Q_i-\sum_{j=1}^N\frac{1}{2}\theta_{ij}P_j,P_i).
\label{i4}
\end{align}
The noncommutative classical Lagrangian can be written in terms of the commutative coordinates and momenta as $P\dot{Q}-H(Q-P/2\theta,~P)$.  
And by using the inverse transformation of (\ref{i2}), we can obtain the noncommutative classical Lagrangian with the noncommutative variables\cite{Girotti}. Then the noncommutative commutation relations are given as the Dirac brackets. 
In the analysis of the noncommutative phase space, the momentum variables are usually
regarded as commutative ones.
In our opinion, however, it is not natural to introduce only coordinate-coordinate noncommutativity.
Because in the Hamiltonian treatment the momenta $p$ are variables independent of and on an equal footing with the coordinates $q$\cite{landau}.

In this paper, we introduce not only coordinate-coordinate noncommutativity, but also momentum-momentum noncommutativity. 
Then, we have a following commutation relations:
\begin{align}
[q_i,~q_j]=i\theta_{ij},\qquad [q_i~,p_j]=i\delta_{ij}, \qquad [p_i,~p_j]=i\eta_{ij},
\label{i5}
\end{align}
where $\theta$ and $\eta$ are $N\times N$ antisymmmetric matrices.
An exact representation of the noncommutative phase space variables (\ref{i5}) by the commutative
ones (\ref{i2}) is presented. Similar transformations
have been proposed by several authors\cite{Demetrian,zhang,Berto,Likang}.
For general antisymmetric matrices $\theta$ and $\eta$, however, those proposed forms make
$[q_i,~p_j]$ non-diagonal which may destroy the canonical cummutation relations. 

In \S 2, first, we study transformation from the commutative phase space coordinates to the noncommutative ones, for $\theta=\eta$. 
And we show that the isotropic case ($\theta=\eta$) can be related to anisoropic case ($\theta\neq\eta$), by using an regular matrix.
As an example, the energy levels of the three-dimensional harmonic oscillator are examined
in \S 3.
The last section is devoted to conclusion.

\section{Transformation}
First, we consider the isotropic noncommutative phase space in which $\theta=\eta=\alpha$
\begin{align}
&[q_i,~q_j]=i\alpha_c\epsilon_{ij},\hspace{0.8em} [q_i~,p_j]=i\delta_{ij},\hspace{0.8em} [p_i,~p_j]=\alpha_c\epsilon_{ij},
\quad(i,j=1,2,\dots,N).
\label{t1}
\end{align}
Here, $\alpha_c$ is a constant, and $\epsilon_{ij}$ is the Levi-Civita symbol. The $N\times N$ antisymmetric matrix $\alpha$ is written as $\alpha\equiv\alpha_c\epsilon_{ij}$.
To obtain the representations of the noncommutative $q_i$ and $p_i$ in term of the corresponding commutative coordinate $Q_i$ and momentum $P_i$, we assume
\begin{align}
\left\{
\begin{aligned}
&q_i=\sum_{j=1}^N(a_{ij}Q_j+b_{ij}P_j),\cr
&p_i=\sum_{j=1}^N(c_{ij}Q_j+d_{ij}P_j),
\end{aligned}
\right.
i=1,\dots,N,
\label{t2}
\end{align}
or, equivalently in the matrix notation,
\begin{align}
\left\{
\begin{aligned}
&q=aQ+bP,\cr
&p=cQ+dP.
\end{aligned}
\right.
\label{t3}
\end{align}
Substituting (\ref{t3}) into (\ref{i5}) and using (\ref{i3}), we obtain conditions
\begin{align}
\begin{aligned}
&ab^{\mathrm{T}}-ba^{\mathrm{T}}=\alpha,\cr
&ad^{\mathrm{T}}-bc^{\mathrm{T}}=1,\cr
&cd^{\mathrm{T}}-dc^{\mathrm{T}}=\alpha,
\end{aligned}
\label{t4}
\end{align}
where $1$ is the identity matrix, the superscript T denotes the matrix transposition.
In order to find the forms of the coefficient matrices $a$, $b$, $c$ and $d$,
let us impose the following ansatz:
\begin{align}
\begin{aligned}
&a=A(\alpha^2), \qquad b=B(\alpha^2)\alpha,\cr
&c=C(\alpha^2)\alpha, \qquad d=D(\alpha^2).
\end{aligned}
\label{t5}
\end{align}
Here $A$ and $B$ ($C$ and $D$) are analytic functions of $\alpha^2$,
whose forms are to be fixed.
For arbitrary functions $f(\alpha^2)$,
we have 
\begin{align}
&f(\alpha^2)\alpha=\alpha f(\alpha).
\label{t6}
\end{align}
These relations can easily be seen by noting that a function regular at $\alpha_c=0$ is a power series of the variable.\footnote{The noncommutativity parameter $\alpha_c$ is assumed to be small\cite{Berto}.
Therefore, we suppose the functions $A$, $B$, $C$ and $D$ are regular at zero.}
By using the relations (\ref{t6}) and noting $[f(\alpha^2)\alpha]^{\mathrm{T}}=-f(\alpha^2)\alpha$,
the conditions (\ref{t4}) reduce to
\begin{align}
\begin{aligned}
-[A(\alpha^2)B(\alpha^2)+B(\alpha^2)A(\alpha^2)]\alpha&=\alpha,\cr
A(\alpha^2)D(\alpha^2)+B(\alpha^2)C(\alpha^2)\alpha^2&=1,\cr
\alpha[C(\alpha^2)D(\alpha^2)+D(\alpha^2)C(\alpha^2)]&=\alpha.
\end{aligned}
\label{t7}
\end{align}
Hence, we have
\begin{align}
&A(\alpha^2)B(\alpha^2)=-\frac{1}{2},
\label{t8-1}\\
&A(\alpha^2)D(\alpha^2)+B(\alpha^2)C(\alpha^2)\alpha^2=1,
\label{t8-2}\\
&C(\alpha^2)D(\alpha^2)=\frac{1}{2}.
\label{t8-3}
\end{align}
Consequently, we obtain
\begin{align}
&A(\alpha^2)B(\alpha^2)=-\frac{1}{2},
\label{t9-1}\\
&A(\alpha^2)D(\alpha^2)=\frac{1\pm\sqrt{1+\alpha^2}}{2},
\label{t9-2}\\
&C(\alpha^2)D(\alpha^2)=\frac{1}{2},
\label{t9-3}\\
&B(\alpha^2)C(\alpha^2)\alpha^2=\frac{1\mp\sqrt{1+\alpha^2}}{2}.
\label{t9-4}
\end{align}
Note that $(1+\sqrt{1\pm\alpha^2})/2$ is the analytic function of the matrix $\alpha^2$.
When $\alpha^2=-1$, we can easily find that this transformation (\ref{t3}) has no inverse transformation.
Thus, we find that the transformation (\ref{t3}) should be of the following form: 
\begin{align}
\left\{
\begin{aligned}
&q=A(\alpha^2)Q-\frac{1}{2}(A(\alpha^2))^{-1}\alpha P,\cr
&p=D(\alpha^2)P+\frac{1}{2}(D(\alpha^2))^{-1}\alpha Q,\cr
&A(\alpha^2)D(\alpha^2)=\frac{1+\sqrt{1+\alpha^2}}{2}.
\end{aligned}
\right.
\label{t10}
\end{align}
The inverse transformation can be easily obtained as
\footnote{$\sqrt{1+\alpha^2}$ is also analytic function of matrix $\alpha$.}
\begin{align}
\left\{
\begin{aligned}
&Q=\frac{1}{\sqrt{1+\alpha^2}}\left(D(\alpha^2)q+\frac{1}{2}(A(\alpha^2))^{-1}\alpha p\right),\cr
&P=\frac{1}{\sqrt{1+\alpha^2}}\left(A(\alpha^2)p-\frac{1}{2}(D(\alpha^2))^{-1}\alpha q\right).
\end{aligned}
\right.
\label{t11}
\end{align}
In this case, we can consider the commutative limit $\alpha_c\rightarrow0$.

In addition, we consider the following transformation 
\begin{align}
\bar{q}\equiv s q,\quad \bar{p}\equiv (s^{-1})^{\mathrm{T}} p
\label{t12}.
\end{align}
Here $s$ is an arbitrary regular matrix which does not satisfy $s^\dagger s=1$.
The variables $\bar{q}$ and $\bar{p}$ satisfy the commutation relations,
\begin{align}
[\bar{q}_i,~\bar{q}_j]=i\theta_{ij},\quad [\bar{q}_i,~\bar{p}_j]=i\delta_{ij},\quad
[\bar{p}_i,~\bar{p}_j]=i\eta_{ij}.
\label{t13}
\end{align}
Here $\theta\equiv s\alpha s^{\mathrm{T}}$ and $\eta\equiv (s^{-1})^{\mathrm{T}}\alpha s^{-1}$ are antisymmetric matrices.
Thus $(\bar{q},~\bar{p})$ also has the noncommutative commutation relations.
Then the variables $Q$ and $P$ transform as follows:
\begin{align}
\bar{Q}\equiv s Q,\quad \bar{P}\equiv (s^{-1})^{\mathrm{T}} P.
\label{t14}
\end{align}
And the transformation (\ref{t10}) can be rewritten in terms of 
$\bar{q}$, $\bar{p}$, $\bar{Q}$, $\bar{P}$, $\theta$ and $\eta$ as follows:
\begin{align}
\left\{
\begin{aligned}
&\bar{q}=A(\theta\eta)\bar{Q}-\frac{1}{2}(A(\theta\eta))^{-1}\theta \bar{P},\cr
&\bar{p}=D(\theta\eta)\bar{P}+\frac{1}{2}(D(\theta\eta))^{-1}\eta \bar{Q},\cr
&A(\theta\eta)D(\theta\eta)=\frac{1+\sqrt{1+\theta\eta}}{2}.
\end{aligned}
\right.
\label{t15}
\end{align}
This expression tell us, we can choose any noncommutative parameters $\eta$ and $\theta$, by using the transformation (\ref{t12}).
In other words, an anisotropic noncommutative phase space can be representated as a rescaling of an arbitrary isotropic noncommutative phase space.
The transformation (\ref{t10}) and (\ref{t15}) are linear, and we can also easily find matrix form of the transformation which corresponds to (\ref{t10}) and (\ref{t15}) by using the regular matrix $s$. 
The antisymmetric matrix $\alpha$ can be block-diagonalized as $\alpha_{\mathrm{BD}}$,  by using an orthogonal matrix $R$. 
When $R$ is taken as $s$, $A(\alpha_{\mathrm{BD}}^2)$ and $D(\alpha_{\mathrm{BD}}^2)$ should be diagonal matrix.
By $R$, the coordinates and momenta are converted into $q_{\mathrm{BD}}$ and $p_{\mathrm{BD}}$.
Thus, in the noncommutative phase space $(q_{\mathrm{BD}},~p_{\mathrm{BD}})$, transformation (\ref{t15}) can be written an exact matrix form. 
For the noncommutative phase space $(q_{\mathrm{BD}},~p_{\mathrm{BD}})$, 
we take the regular matrix $s$ as a product of a diagonal matrix and the orthogonal matrix $R^{\mathrm{T}}$,
then one can find an exact matrix form of (\ref{t15}).

For example, we take the matrix $s$ as 
$\diag(\frac{\beta_1}{\sqrt{\alpha_c}},\frac{\beta_2}{\sqrt{\alpha_c}},\dots,\frac{\beta_N}{\sqrt{\alpha_c}})$.
Here $\alpha_c$ and $\beta_i$ are constants.
Then the variables $\bar{q}$ and $\bar{p}$ satisfy the commutation relations (\ref{t13}).
And the noncommutative parameters are defined as $\theta_{ij}\equiv\epsilon_{ij}\beta_i\beta_j$ and $\eta_{ij}\equiv\epsilon_{ij}\frac{\alpha_c^2}{\beta_i\beta_j}$(repeated indices are not summation convention), 
(\ref{t10}) can be rewritten as follows,
\begin{align}
\left\{
\begin{aligned}
&\bar{q}=A(\alpha^2)\bar{Q}-\frac{1}{2}(A(\alpha^2))^{-1}\theta \bar{P},\cr
&\bar{p}=D(\alpha^2)\bar{P}+\frac{1}{2}(D(\alpha^2))^{-1}\eta \bar{Q},\cr
&A(\alpha^2)D(\alpha^2)=\frac{1+\sqrt{1+\alpha^2}}{2}.
\end{aligned}
\right.
\label{t16}
\end{align}
Thus, this phase space is anisotropically noncommutative.
In the phase space $(\bar{q},~\bar{p})$, we can consider 
$\alpha_c\rightarrow0$ limit. Then the commutation relations (\ref{t13}) turns to  
\begin{align}
&[\bar{q}_i,~\bar{q}_j]=i\theta_{ij},\qquad [\bar{q}_i~,\bar{p}_j]=i\delta_{ij},\qquad [\bar{p}_i,~\bar{p}_j]=0.
\label{t17}
\end{align}
Furthermore, when we choose $\diag^{-1}(\frac{\beta_1}{\sqrt{\alpha_c}},\frac{\beta_2}{\sqrt{\alpha_c}},\dots,\frac{\beta_N}{\sqrt{\alpha_c}})$ as the regular matrix $s$, 
then $\theta_{ij}$ and $\eta_{ij}$ are defined as $\epsilon_{ij}\frac{\alpha^2_c}{\beta_i\beta_j}$ and
$\epsilon_{ij}\beta_i\beta_j$.
And we can consider $\alpha_c\rightarrow0$ limit
in the phase space $(\bar{q},~\bar{p})$, then the commutation relations (\ref{t13}) can be rewritten as follows:
\begin{align}
&[\bar{q}_i,~\bar{q}_j]=0,\qquad [\bar{q}_i~,\bar{p}_j]=i\delta_{ij},\qquad [\bar{p}_i,~\bar{p}_j]=i\eta_{ij}.
\label{t18}
\end{align}
Note that, in above two cases,
the commutative limit should be taken $\beta_i\beta_j\rightarrow0$ after the limit $\alpha_c\rightarrow0$.

Through these resutlts,
while keeping the commutation relation between coordinate and momentum,
increasing of coordinate and decreasing of momentum (or conversely decreasing coordinate and increasing mommentum ) is changing the noncommutative parameter $\alpha$.
In other words, an arbitrary noncommutative phase space can be tranformed into the other noncommutative phase space by rescaling the noncommutative variables keeping the commutation relation between the coordinates and the momenta.
Thus, we can regard an isotropic noncommutative quantum mechanics which has the commutation relation (\ref{t1}) as a basis.
\section{Three-demensional Harmonic oscillator}
In this section, we consider the three-dimensional harmonic oscillator in the noncommutative phase space.
Under the condition $A(\alpha^2)=D(\alpha^2)$, the transformation (\ref{t10}) is rewritten as  
\begin{align}
\left\{
\begin{aligned}
&q=x_+(\alpha^2)Q-\frac{1}{2}(x_+(\alpha^2))^{-1}\alpha P,\cr
&p=x_+(\alpha^2)P+\frac{1}{2}(x_+(\alpha^2))^{-1}\alpha Q,\cr
&x_+(\alpha^2)=\sqrt{1+\sqrt{1+\alpha^2}}.
\end{aligned}
\right.
\label{ho1}
\end{align}
In the following, we adopt this expression instead of (\ref{t10}), and we apply it to the three-dimensional harmonic oscillator in the noncommutative phase space.

The Hamiltonian for the three-dimensional harmonic oscillator is given by
\begin{align}
\mathscr{H}=\frac{1}{2M}\sum_{i=1}^3\left(p_ip_i+M^2\omega^2q_iq_i\right).
\label{ho2}
\end{align}
First, we consider the isotropic noncommutative phase space which has the commutation relation
\begin{align}
[q_i,~q_j]=i\alpha_c\epsilon_{ij},
~[q_i,~p_j]=i\delta_{ij},~
[p_i,~p_j]=i\alpha_c\epsilon_{ij}\quad(i=1,2,3).
\label{ho3}
\end{align}
The antisymmetric matrix $\alpha$ can be block-diagonalized by using the orthogonal matrix $R$.
After block-diagonalization of $\alpha$, 
\begin{align}
\begin{aligned}
&[\bar{q}_a,~\bar{q}_b]=i\bar{\alpha}\epsilon_{ab},~[\bar{q}_a,~\bar{p}_b]=i\delta_{ab},~
[\bar{p}_a,~\bar{p}_b]=i\bar{\alpha}\epsilon_{ab},\cr
&[ \bar{q}_a,~\bar{q}_3]=0,~[\bar{q}_3,~\bar{p}_3]=i,~[\bar{p}_a,~\bar{p}_3]=0,
\end{aligned}
\quad(a,b=1,2).
\label{ho4}
\end{align}
And we take the regular matrix $s$ as
\begin{align}
s\equiv \diag\left(\sqrt{\frac{M\omega\beta}{\bar{\alpha}}},\sqrt{\frac{M\omega\beta}{\bar{\alpha}}},\sqrt{\frac{M\omega\beta}{\bar{\alpha}}}\right)R,
\label{ho5}
\end{align}
here $\bar{\alpha}$ and $\beta$ are constants.
This case corresponds to that $\theta$ and $\eta$ are simultaneously diagonalizable.
Then the Hamiltonian (\ref{ho2}) is rewritten as
\begin{align}
\mathscr{H}=&\frac{1}{2M}\sum_i^3\left(\bar{p}_i\bar{p}_i+M^2\omega^2\bar{q}_i\bar{q}_i\right)\quad(i=1,2,3),\cr
=&\frac{M\omega}{2}\chi\left(A^\dagger_+A_++A^\dagger_-A_-
+\frac{1}{2}\varTheta(A^\dagger_+A_+-A^\dagger_-A_-)+1\right)\cr
&+\frac{M\omega}{2}(2A^\dagger_3A_3+1),
\label{ho6}
\end{align}
where 
\begin{align}
\begin{aligned}
\chi&\equiv\frac{\sqrt{\left(\beta^2 M^2\omega^2+x_+^4\right)\left(\bar{\alpha}^4+x_+^4\beta^2 M^2 \omega^2\right)}}{2 \beta  M x_+^2 \omega},\cr
\varTheta&\equiv\frac{x_+^2 \left(\bar{\alpha}^2+\beta^2 M^2 \omega^2\right)}{\sqrt{\left(\beta^2 M^2 \omega^2+x_+^4\right) \left(\bar{\alpha}^4+\beta^2 M^2 x_+^4 \omega^2\right)}},
\end{aligned}
\label{ho7}
\end{align}
and the creation and annihilation operators $A_\rho^\dagger$, $A_\rho$ are defined as 
\begin{align}
\begin{aligned}
&A^\dagger_\pm\equiv\frac{1}{\sqrt{2}}(a^\dagger_1\pm i a^\dagger_2),\quad A^\dagger_3\equiv a^\dagger_3,\cr
&A_\pm\equiv\frac{1}{\sqrt{2}}(a_1\mp i a_2),\quad A_3\equiv a_3.
\end{aligned}
\label{ho8}
\end{align}
Here $a_i^\dagger$ and $a_i$ are also creation and annihilation operators respectively.
They can be written in terms of $\bar{Q}$ and $\bar{P}$ as
\begin{align}
\begin{aligned}
a^\dagger_i\equiv\frac{1}{\sqrt{2}}\left(\frac{1}{\xi_i}\bar{Q}_i-i\xi_i\bar{P}_i\right),\cr
a_i\equiv\frac{1}{\sqrt{2}}\left(\frac{1}{\xi_i}\bar{Q}_i+i\xi_i\bar{P}_i\right),
\end{aligned}
\label{ho9}
\end{align}
where $\xi_i$ have following forms: 
\begin{align}
\xi_1=\xi_2\equiv\sqrt{\beta M \omega} \sqrt[4]{\frac{\beta^2 M^2 \omega^2+x_+^4}{\bar{\alpha}^4+\beta^2 M^2 x_+^4 \omega^2}},
\quad
\xi_3\equiv\sqrt{M\omega}.
\label{ho10}
\end{align}
$A_\rho$ and $a_i$ satisfy the commutator algebra
\begin{align}
&[a_i,~a_j]=0,\quad [a_i,~a_j^\dagger]=\delta_{ij},\quad
[a_i^\dagger,~a_j^\dagger]=0,\quad(i,j=1,2,3)
\label{ho11-1}\\
&[A_\rho,~A_\sigma]=0,\quad
[A_\rho,~A^\dagger_\sigma]=\delta_{\rho\sigma},\quad
[A^\dagger_\rho,~A^\dagger_\sigma]=0.\quad
\quad(\rho,\sigma=\pm,3)
\label{ho11-2}
\end{align}
In the sequel, we obatain the Schr\"{o}dinger equation in the noncommutative phase space in terms of the commutative variables
\begin{align}
H|n_\pm,n_3\rangle=\frac{M\omega}{2}\left[\chi\left(N_++N_-+\frac{1}{2}\varTheta(N_+-N_-)+1\right)+(2N_3+1)\right]|n_\pm,n_3\rangle,
\label{ho12}
\end{align}
where $N_\pm\equiv A^\dagger_\pm A_\pm$ and $N_3\equiv A^\dagger_3A_3$.
Since $[N_\rho,~N_\sigma]=0$,
the eigenvector $|n_\pm,n_3\rangle$ is given by
\begin{align}
|n_\pm,n_3\rangle=\frac{1}{\sqrt{N_+!}}\frac{1}{\sqrt{N_-!}}\frac{1}{\sqrt{N_3!}}
\left(A^\dagger_+\right)^{n_+}\left(A^\dagger_-\right)^{n_-}\left(A^\dagger_3\right)^{n_3}|0,0,0\rangle .
\label{ho13}
\end{align}

Futhermore, it is useful to recognize that the system under consideration possesses, its commutative couterpart as well as, the $SU(2)$ symmetry whose generators ($J_1$, $J_2$, $J_3$) and the Casimir operator ($J^2$) are
\begin{align}
J_1&=\frac{1}{2}(a^\dagger_1a_2+a_1a^\dagger_2),
\label{ho14-1}\\
J_2&=\frac{i}{2}(a^\dagger_1a_2-a_1a^\dagger_2),
\label{h014-2}\\
J_3&=\frac{1}{2}(a^\dagger_1a_1-a^\dagger_2a_2),
\label{ho14-3}\\
J^2&=\sum_k^3J_kJ_k=\frac{N}{2}\left(\frac{N}{2}+1\right),
\label{ho14-4}
\end{align}
where $N=a^\dagger_1a_1+a^\dagger_2a_2$. 
We can easily find $N=N_++N_-$ and $J_2=\frac{1}{2}(N_+-N_-)$.
It is straightforward to verify that (\ref{ho14-1})-(\ref{ho14-4}) implies $[J_k , J_l] = i \epsilon_{klm} J_m$ and $[J^2 , J_k] = 0$, as it must be. Since $[N , J_k] = 0$, it follows from (\ref{ho12}) that the energy eigenvalue problem leads
\begin{align}
H|j,m,n_3\rangle=\frac{M\omega}{2}\left[\chi\left(N+2\varTheta J_2 +1\right)+(2N_3+1)\right]|j,m,n_3\rangle
\label{ho15}
\end{align}
where $|j , m\rangle$ is the common eigenvector of $J^2$ and $J_2$, and $|j,m,n_3\rangle=|n_\pm,n_3\rangle$. They are labeled by the eigenvalues of $J^2$ and $J_2$, named $j$ and $m$, respectively. As is well known, 

\begin{align}
\begin{aligned}
&j = 0, 1/2, 1, 3/2, ...,\cr
&-j \leq m \leq j,
\end{aligned}
\label{ho16}
\end{align}
while $n=2j$.
$\varTheta$ is not equal to zero for any real number $\theta$ and $\eta$.
Thus the angular monmentum term $J_2$ is always appear in the noncommutative harmonic oscillator, and it can have the half integer eigenvalue.
It is easy to find that this term $J_2$ is also appear in a pair of harmonic oscillators $A^\dagger_+ A_+$ and $A^\dagger_- A_-$.
Certainly, this angular momentum term also appears when we treate the noncommutative harmonic oscillator which has only coordinate-coordinate noncommutativity in the Moyal product method\cite{Hatzikintas,kumar}.
Naturaly, this angular momentum term which is appear in the pair of harmonic oscillator is canceled in the commutative limit\cite{chda,duja,mopo,horv}.

Next, we comment on that $\theta=s\alpha s^\mathrm{T}$ and $\eta=(s^{-1})^\mathrm{T}\alpha s^{-1}$ aren't simultaneously block-diagonalizable. Then the regular matrix $s$ can be given 
\begin{align}
s\equiv \diag\left(\sqrt{\frac{M\omega\beta}{\bar{\alpha}}(1+\varepsilon_1)},\sqrt{\frac{M\omega\beta}{\bar{\alpha}}(1+\varepsilon_2)},\sqrt{\frac{M\omega\beta}{\alpha
_c}(1+\varepsilon_3)}\right)
R,
\label{ho17}
\end{align}
where $\bar{\alpha}$, $\beta$ and $\varepsilon_{i}$ are constants. $\varepsilon_i$ are shifts of the noncommutative parameters from the antisymmetric matrices $\theta$ and $\eta$ are simultaneously diagonalizable case. The noncommutative parameter is considered very small, so $\varepsilon_i$ is also very small.
The Hamiltonian is 
\begin{align}
{\mathscr{H}}=H+H',
\label{ho18}
\end{align}
where $H$ is (\ref{ho12}), and $H'$ is defined as
\begin{align}
H'=&~
{\cal{O}}_1(\varepsilon_i)A^\dagger_+ A_++{\cal{O}}_2(\varepsilon_i)A^\dagger_- A_-
+{\cal{O}}_3(\varepsilon_i)A^\dagger_3A_++{\cal{O}}_4(\varepsilon_i)A^\dagger_+ A_3\cr
&+{\cal{O}}_5(\varepsilon_i)A^\dagger_3A_-+{\cal{O}}_6(\varepsilon_i)A^\dagger_- A_3
+{\cal{O}}_7(\varepsilon_i)A_+ A_-+{\cal{O}}_8(\varepsilon_i)A^\dagger_+ A^\dagger_-\cr
&+{\cal{O}}_9(\varepsilon_i)A_- A_++{\cal{O}}_{10}(\varepsilon_i)A^\dagger_- A^\dagger_+
+{\cal{O}}_{11}(\varepsilon_i)A_3A_++{\cal{O}}_{12}(\varepsilon_i)A^\dagger_3A^\dagger_+\cr
&+{\cal{O}}_{13}(\varepsilon_i)A_3A_-+{\cal{O}}_{14}(\varepsilon_i)A^\dagger_3A^\dagger_-.
\label{ho19}
\end{align}
In this case, the second term $H'$ may be treated as a perturbation.
We give that the three dimensional harmonic oscillator in the noncommutative phase space which the coordinate and momentum satisfy the commutation relation (\ref{i5}) may calculate as a pertabative method for the arbitrary noncommutative parameter $\theta$ and $\eta$.
\section{Conclusion}
We have obtained the exact linear transformation in the phase space which relates an arbitrary noncommutative 
phase space to a commutative one.
Using a regular matrix, an arbitrary isotropic noncommutative parameter is converted into any anisotropic noncommutative parameter (or conversely an arbitrary anisotropic noncommutative parameter is converted into the isotrocpic noncommutative parameter).
And we can obtain an exact matrix form of transfromation by using suitable regular matirx.
Thus, we can consider (\ref{t10}) as a basis of the transformation.
It is a special case that either coordinates or momenta are noncommutative and the other are commutative.

The energy of the three-dimensional harmonic oscillator described by the noncommutative coordinate and momentum variables was calculated as an example.
When the noncommutative parameter $\theta$ and $\eta$ are simultaneously diagonalizable, the three dimensional harmonic oscillator is described by the one-dimensional harmonic oscillator and the two-dimensional noncommutative harmonic oscillator.
Except for the commutative limit, a pair of harmonic oscillators with the angular momentum terms always appear.
The angular momentum terms with oppsite sign each other, which can have a half integer eigenvalue, and those are canceled as a result of the commutative limit.
In the case of that $\theta$ and $\eta$ aren't simultaneously diagonalizable, interaction term of the two-dimensional noncommutative harmonic oscillator and the commutative oscillator appears in the Hamiltonian as a pertabation.

We expect that the $N$-dimensional harmonic oscillator in the noncommutative phase space represents in a spinor framework, because the angular momentum term has the half integer eigenvalue. 
\section*{Acknowledgments}
We would like to thank Shinji Hamamoto, Minoru Hirayama and Takenori Fujikawa for many useful discussions and 
reading the manuscript.

\end{document}